\documentclass[aps,prb,twocolumn,groupedaddress,showpacs,showkeys]{revtex4}

\usepackage{xspace}
\usepackage{epsfig}
\usepackage{amsfonts}

\newcommand{\fig}[4]{
\begin{figure}[#1]
\begin{center}
\includegraphics [width=#2 in, keepaspectratio=true] {#3.eps}
\caption{#4 \label{fig:#3}}
\end{center}
\end{figure}}

\begin{document}


\title{On the quantum and classical scattering times due to
charged dislocations in an impure electron gas }

\author{Debdeep Jena}
\email[Electronic mail: ]{djena@engineering.ucsb.edu}
\author{Umesh K. Mishra}
\affiliation{Department of Electrical and Computer Engineering and Materials Department \\
            University of California, Santa Barbara \\
            CA, 93106}

\date{\today}

\begin{abstract}
We derive the ratio of transport and single particle relaxation
times in three and two - dimensional electron gases due to
scattering from charged dislocations in semiconductors.  The
results are compared to the respective relaxation times due to
randomly placed charged impurities.  We find that the ratio is
larger than the case of ionized impurity scattering in both three
and two-dimensional electron transport.
\end{abstract}


\maketitle

%

In their work on the transport properties of impure metallic
systems, Das Sarma and Stern \cite{dss} show the difference
between two characteristic relaxation times for the case of
scattering from randomly located charged impurities.  They make a
clear distinction between the transport scattering time $\tau_{t}$
and the quantum scattering time (or the single particle relaxation
time) $\tau_{q}$.

Whilst the quantum scattering time is a measure of the time spent
by a carrier in a particular momentum eigenstate state $| {\bf
k}\rangle$ in the presence of a perturbing (or scattering)
potential, the transport scattering time is the measure of the
time spent in {\em moving along the applied field}\cite{hsu_walu};
the lifetime thus gets weighted by an angular contribution factor
$(1-\cos\theta)$ that enhances small angle scattering contribution
to the relaxation rate over large angle scattering. Thus, the
scattering rate is defined to be

\begin{equation}
\frac{1}{\tau}=\sum_{k'}S({\bf k',k})f(\theta)
\end{equation}

where $f(\theta)=1$ or $(1-\cos\theta)$ for quantum and transport
relaxation rates respectively.  Using this result, Das Sarma and
Stern have shown that the ratio $\tau_{t}/\tau_{q}$ exceeds unity
for ionized impurity scattering for both the three and
two-dimensional metallic electron gases.

Among the elastic scattering processes, the ratio deviates
strongly from unity for {\em Coulombic} scatterers where the
scattering potential has a long-range nature.  Short range
scattering processes such as due to alloy scattering are
isotropic, and so are inelastic scattering processes such as due
to phonons.  For such scatterers, the ratio remains close to
unity.  Gold\cite{gold} extended the results of Das-Sarma and
Stern to include interface roughness and alloy scattering in
two-dimensions.  The two main Coulombic scattering mechanisms in
semiconductors originate from randomly distributed individual
charge centers (such as those from dopants, impurities, or charged
vacancies) or from charged dislocations.

The problem of the ratio of transport to quantum scattering times
has not yet been solved for dislocation scattering.  Scattering
from charged dislocations has become an increasingly important
topic owing to its relevance in III-V nitride semiconductor
structures.  Look\cite{look_prl} recently showed the importance of
charge dislocation scattering in bulk GaN.  The effect of
dislocation scattering on two-dimensional electron gas
conductivity was also studied\cite{myfirstapl}.  Both the works
calculate the transport scattering times to see its effect on
electron mobilities.  Many workers have reported the transport to
quantum scattering time ratios for AlGaN/GaN two-dimensional
electron gases(2DEGs)\cite{elhamri, arakawa, frayssinet, saxler,
elhamriJAP, brana}.  These 2DEGs typically have a large number of
dislocations whose effect on the quantum scattering time has not
yet been studied.  The recent demonstration\cite{MyNewAPL} of
polarization-doped degenerate three-dimensional electron gases in
graded gap AlGaN/GaN structures presents another experimental
system which requires a study of the ratio $\tau_{t}/\tau_{q}$ for
the identification of major scattering mechanisms.

In this work, we present the theory of quantum scattering times
for charged dislocation scattering for both three and
two-dimensional electron gases.  First, we derive a closed form
expression for the ratio $\tau_{t}/\tau_{q}$ for dislocation
scattering in the three-dimensional electron gas (3DEG). We
compare the results to the case of ionized impurity scattering. We
then derive the ratio for the two-dimensional electron gas(2DEG)
and compare the ratio to remote ionized impurity scattering.

We now derive an expression for the quantum scattering time of
electrons for three-dimensional carriers for scattering from
charged dislocations.  We assume that there are $N_{disl}$
parallel dislocations per unit area that pierce a
three-dimensional electron gas of density $n$ along the $z$
direction.  We set $c$ to be the distance between the individual
charges on the dislocation, making it a line charge of density
$1/c$.

The screened Coulombic potential due to such a charged dislocation
immersed in the gas of mobile carriers is given by\cite{seeger}

\begin{equation}
V_{disl}^{scr}(r)=\frac{e}{2 \pi \epsilon
c}K_{0}(\frac{r}{\lambda})
\end{equation}

where $K_{0}$ is a zero-order modified Bessel function, and
$\lambda$ is the screening length.  To find the matrix element
$\langle {\bf k'}|V_{disl}^{scr}(r)| {\bf k }\rangle$ for
scattering, we note the fact that the scattering is
two-dimensional, affecting only the component of momentum of
incident electrons perpendicular to the dislocation axis, ${\bf
k_{\perp} }$.  Under these assumptions, the dislocation scattering
matrix rate can be shown\cite{look_prl} to be given by

\begin{equation}
S_{disl}(k',k)=\frac{2 \pi}{\hbar}(\frac{e^{2}}{\epsilon c
}\frac{\lambda^{2}}{[1+(q\lambda)^{2}]})^{2}\delta(E_{k_{\perp}}-E_{k_{\perp}'})
\end{equation}

where $q^{2}=|{\bf k_{\perp}^{'} - k_{\perp}
}|^{2}=2k_{\perp}^{2}(1-\cos\theta)$.


\fig{t}{2.9}{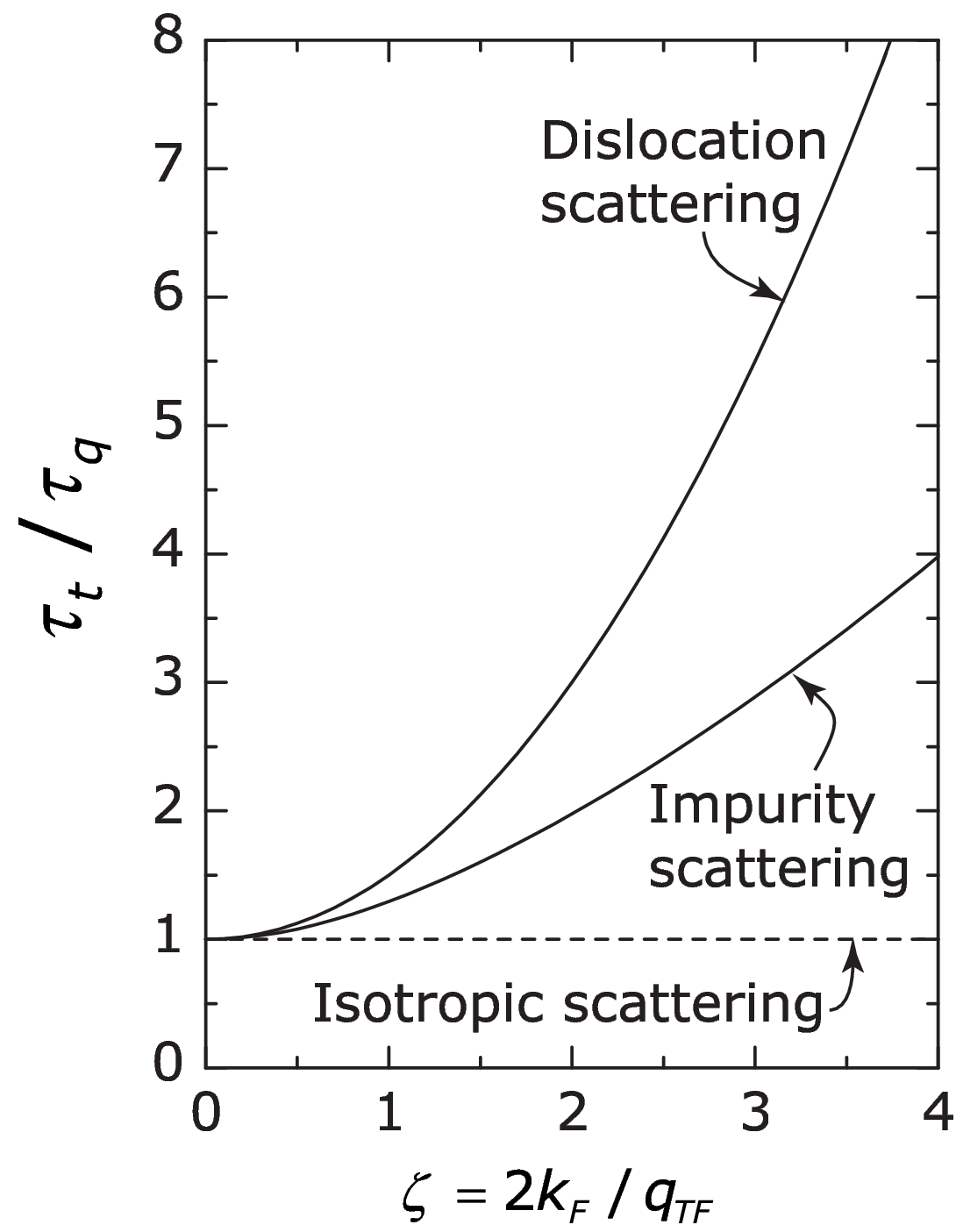}{Plot showing the lifetime ratio for
scattering from charged dislocations and ionized impurities for a
degenerate electron gas as a function of the carrier density.  The
ratio deviates significantly for the two Coulombic scattering
processes from the normal value of unity for all other
non-Coulombic isotropic scattering processes.  The plot is
generated from the {\em exact} relations for the lifetime ratio
for the two scattering mechanisms derived in this work. }


To find the quantum scattering rate, we have to sum this
scattering rate for all values of $k_{\perp}^{'}$ {\em without}
the $(1-\cos\theta)$ term that is required for calculating the
classical mobility.  Summing this expression over all values of
${\bf k_{\perp}^{'}}$ using the prescription for 2-dimensional DOS
$\sum_{k_{\perp}}(...) \rightarrow 1/(2\pi)^{2}\int
d^{2}k_{\perp}(...) $, and evaluating the integral {\em exactly},
we get the quantum scattering rate due to charged dislocations to
be

\begin{equation}
\frac{1}{\tau_{disl}^{q}(k)} =
\frac{N_{disl}e^{4}m^{\star}}{\hbar^{3} \epsilon^{2} c^{2}}
\frac{\lambda^{4}}{(1+4k^{2}\lambda^{2})^{3/2}}\times [
1+2(k\lambda)^{2}].
\end{equation}

Comparing this to the transport scattering rate due to charged
dislocations derived by Podor\cite{podor},

\begin{equation}
\frac{1}{\tau_{disl}^{t}(k)} =
\frac{N_{disl}e^{4}m^{\star}}{\hbar^{3} \epsilon^{2} c^{2}}
\frac{\lambda^{4}}{(1+4k^{2}\lambda^{2})^{3/2}}
\end{equation}

we find the simple relation

\begin{equation}
\frac{\tau_{t}}{\tau_{q}}|_{disl}=1+\frac{1}{2}\zeta^{2}.
\end{equation}

where $\zeta=k\lambda$.  For a metallic 3DEG, transport occurs
between carriers at the Fermi level.  Then, the scattering times
can be evaluated at the Fermi energy, and $\zeta =
2k_{F}\lambda_{TF}$, where $k_{F}=(3\pi^{2}n)^{1/3}$ is the Fermi
wavevector and $\lambda_{TF}=q_{TF}^{-1}=\sqrt{2\epsilon
\varepsilon_{F}/3e^{2}n}$ is the Thomas-Fermi screening length,
$\varepsilon_{F}=\hbar^{2}k_{F}^{2}/2m^{\star}$ being the Fermi
energy.  Under these conditions, the lifetime ratio is dependent
{\em only} on the carrier concentration $n$.

In Figure 1, we plot the ratio of transport and quantum scattering
times against the dimensionless quantity $\zeta$.  In the figure,
we also plot the ratio for random impurity scattering\cite{dss} to
show the relative intensity.  It is seen that the ratio is more
for dislocation scattering than for impurity scattering since
dislocation scattering is {\em inherently} more anisotropic than
impurity scattering.  While impurity scattering potential of point
charges possesses spherical symmetry, dislocation scattering
potential of a line charge is cylindrically symmetric, thus
causing an additional anisotropy in scattering of
three-dimensional carriers. The ratio approaches unity (as is also
true for impurity scattering) as $\zeta \rightarrow 0$.


\fig{t}{2.9}{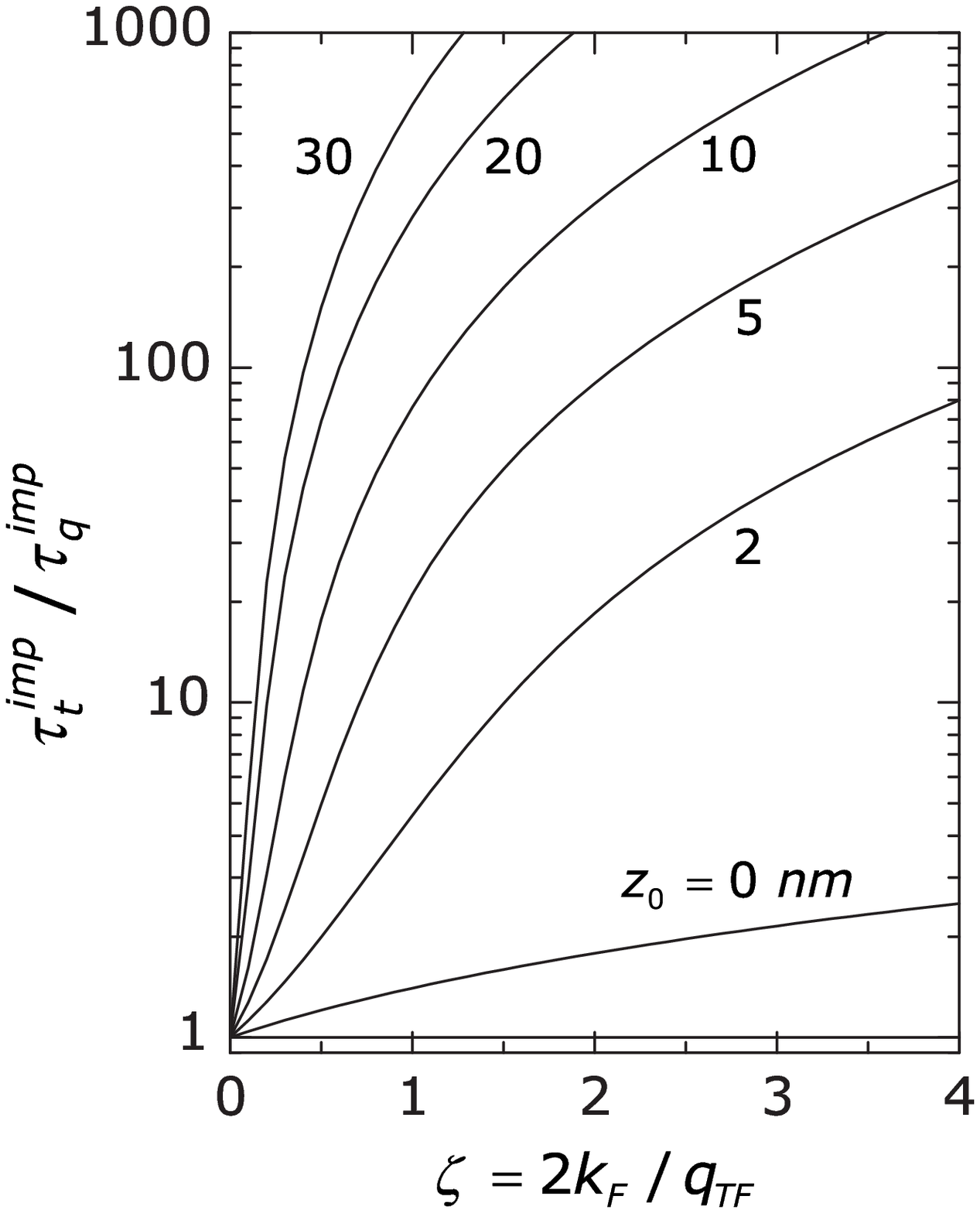}{Plot showing $\tau_{t}/\tau_{q}$
for scattering from remote impurities for an AlGaN/GaN
two-dimensional electron gas.  The different curves are for
changing thickness of the AlGaN barrier.  The ratio increases very
fast with the barrier thickness since remote impurity scattering
strongly enhances small angle scattering, increasing the transport
scattering time.  }


The transport to quantum scattering times ratio due to ionized
impurities for a 2DEG has been studied in some detail in the light
of experimental evidence for both Si-MOSFET inversion layers and
AlGaAs/GaAs modulation-doped 2DEGs.  Das Sarma and Stern show in
their work\cite{dss} how remote ionized impurity scattering causes
the ratio to become very large as the remote donors are placed
farther away from the 2DEG channel, strongly enhancing forward
angle scattering. In a Si-MOSFET, the 2DEG is formed by inversion
induced by a gate voltage.  In a AlGaAs/GaAs heterojunction, the
2DEG is formed by intentional modulation doping from remote
donors.

An AlGaN/GaN 2DEG is distinct from these cases, where the 2DEG
forms due to the strong internal polarization
fields\cite{ambacher} that extract carriers from remote surface
states\cite{ibbo}.  There is no intentional doping for these
structures, and scattering originates from the donor-like surface
states that supply the 2DEG electrons. Thus, the distance of the
remote donors from the 2DEG is fixed by the AlGaN layer thickness.
We calculate the relaxation rate ratio for remote ionized impurity
scattering for such a structure with a changing AlGaN layer
thickness; the results are shown in Figure 2.  As is easily seen,
the ratio becomes much larger than unity as the AlGaN layer
thickness increases.  We will shortly compare this result with the
case of charged dislocation scattering.


\fig{t}{2.9}{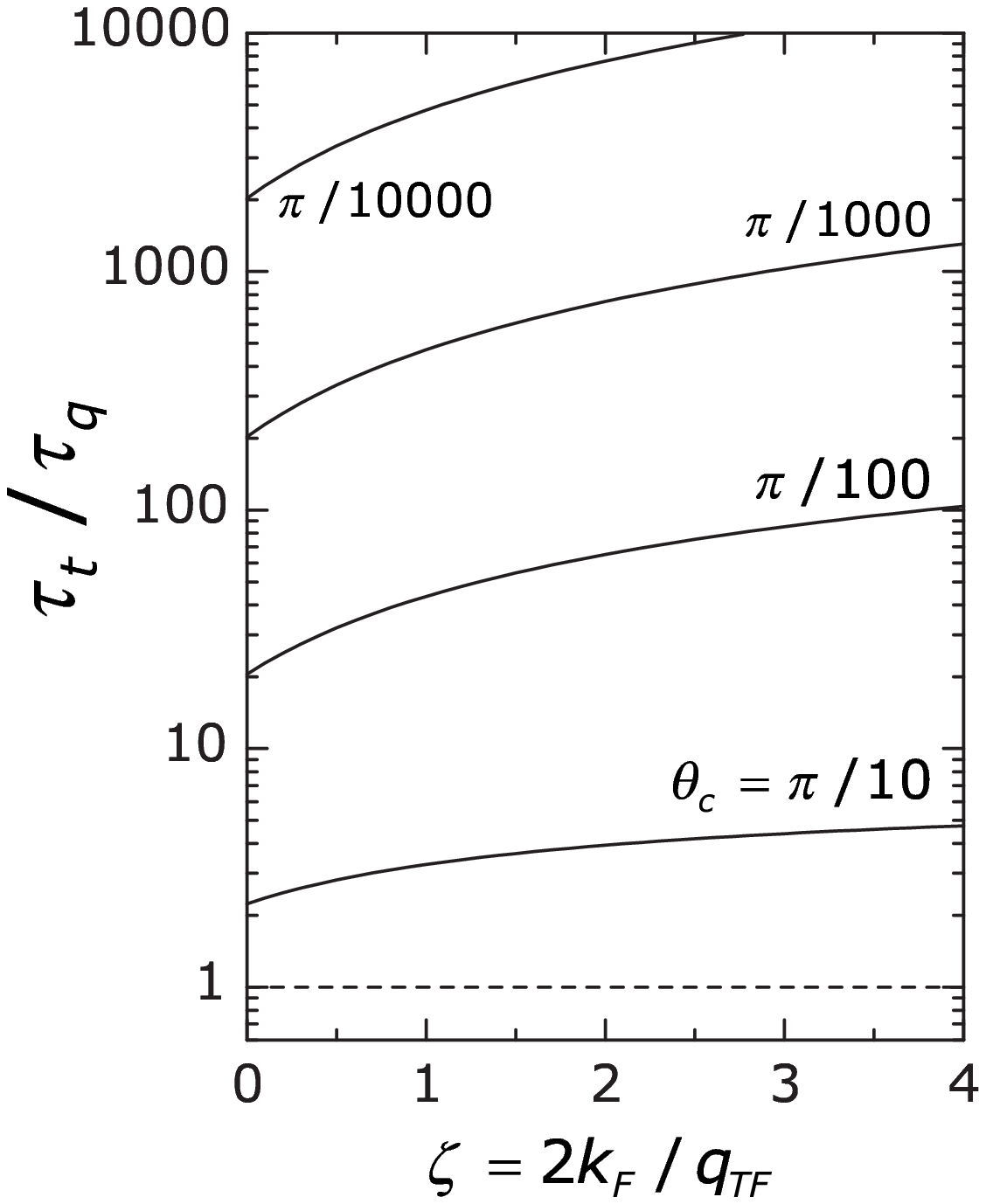}{Plot showing
$\tau_{t}/\tau_{q}$ for scattering from charged dislocations for a
2DEG.  The different curves are for different small angle cutoffs
$\theta_{c}$ for scattering as treated in the paper.  As more
small angle scattering is included, the ratio becomes very large
and eventually diverges as $\theta_{c} \rightarrow 0$.  The
dependence of $\tau_{t}/tau_{q}$ is much weaker for dislocation
scattering than impurity scattering for 2DEGs. }


The transport scattering rate due to charged dislocation
scattering for 2DEG carriers was derived
recently\cite{myfirstapl}.  The screened matrix element for
charged dislocation scattering was derived to be

\begin{equation}
\langle k' | V(r) | k \rangle = \frac{e}{\epsilon c}
\frac{1}{q(q+q_{TF})}
\end{equation}

where $q_{TF}$ is the two-dimensional Thomas-Fermi screening
wavevector and $q=|{\bf k' - k}|=2k_{F}\sin(\theta/2)$ is the
change in the 2D wavevector due to scattering. Using this result,
we derive the quantum scattering rate to be

\begin{equation}
\frac{1}{\tau_{q}}=\frac{N_{dis}m^{\star}e^{2}}{\hbar^{3}\epsilon^{2}c^{2}}
\frac{I_{q}}{4\pi k_{F}^{4}}
\end{equation}

where $I_{q}$, an integral dependent on the dimensionless
parameters $\zeta$ and $u=q/2k_{F}=\sin(\theta/2)$ ($\theta$ is
the angle of scattering) is given by

\begin{equation}
I_{q}=\frac{1}{2}\zeta^{2}\int_{0}^{1} du
\frac{1}{u^{2}(1+\zeta^{2}u^{2})\sqrt{1-u^{2}}}.
\end{equation}

The transport scattering rate is given by

\begin{equation}
\frac{1}{\tau_{t}}=\frac{N_{dis}m^{\star}e^{2}}{\hbar^{3}\epsilon^{2}c^{2}}
\frac{I_{t}}{4\pi k_{F}^{4}}
\end{equation}

where $I_{t}$ is given by

\begin{equation}
I_{t}=\zeta^{2}\int_{0}^{1} du
\frac{1}{(1+\zeta^{2}u^{2})\sqrt{1-u^{2}}}.
\end{equation}

Whilst $I_{t}$ can be evaluated exactly, it can be easily seen
that $I_{q}$ {\em diverges} as $u = \sin(\theta/2) \rightarrow 0$,
or, in other words, when $\theta \rightarrow 0$.  This is the case
of scattering that is strongly peaked in the forward direction.
The ratio of the quantum and transport scattering times, given by
the ratio $\tau_{t}/\tau_{q}$ thus acquires a singularity at small
scattering angles.  We define a scattering angle cutoff
$\theta_{c}$ and evaluate the ratio $\tau_{t}/\tau_{q}$ by
including dislocation scattering restricted to $\theta \geq
\theta_{c}$ only.  We evaluate the ratio for the cutoffs
$\theta_{c}=\pi/10,\pi/100,\pi/1000,\pi/10000$. The results are
shown in Figure 3.

As more small angle scattering contribution is included, the ratio
becomes much larger than unity.  On the other hand, as the
scattering is restricted to be more large angle, the ratio
approaches unity since it mimics isotropic scattering.  Two points
can be made about the results.  First, that the dependence of the
ratio on the 2DEG density is much weaker than for impurity
scattering (Figure II).  Second, that the ratio is much larger
than that for impurity scattering as $\theta_{c} \rightarrow 0$.

One can argue that as opposed to the case of impurity scattering
for 2DEGs where the modulation dopants can be placed at any
distance from the 2DEG, a charged dislocation line {\em always}
has a strong {\em remote impurity} nature due to the geometry.
This causes a strong preference for small angle scattering,
causing the ratio $\tau_{t}/\tau_{q} \rightarrow \infty$ as
$\theta_{c}\rightarrow 0$.  Our model of dislocation scattering
implies that quantum scattering from dislocations should have a
{\em minimal effect} on the broadening of Landau levels which in
all likelihood will be determined by other scattering events.  We
point out here that such a divergence of the quantum scattering
time for a 2DEG was also observed by Gold\cite{gold} for the case
of residual impurity scattering.  In his work, by considering
multiple scattering events contribution to the self energy in the
Green's function, he was able to arrive at a renormalized quantum
scattering time which removed the singularity.  This kind of a
treatment for dislocation scattering in 2DEGs might be an
interesting application of many body theoretical techniques, which
we do not attempt here.

Finally, we mention that the scattering time ratio is independent
of the density of scatters.  However, if one measures $\tau_{t}$
from the mobility and $\tau_{q}$ from the amplitudes of Shubnikov
de-Haas oscillations, one measures the scattering time determined
by the dominant scattering mechanism.  That of course depends on
the density of the scatterers.  Hsu and Walukiewicz have recently
shown that the ratio $\tau_{t}/tau_{q}$ for AlGaN/GaN 2DEGs is the
largest when {\em both} long and short range scattering mechanisms
contribute to the total scattering. The analysis of experimentally
determined $\tau_{t}/\tau_{q}$ ratios in AlGaN/GaN 2DEGs needs to
include the effect of a {\em large} increase in the ratio due to
the effects of dislocation scattering.  The recently demonstrated
polarization-doped three-dimensional electron gases\cite{MyNewAPL}
provide an ideal testing ground for applying our results for the
3DEG case.  By exploiting the internal polarization fields, wide
degenerate electron slabs of high mobility were demonstrated in
graded AlGaN material system.  The 3DEG densities in such electron
slabs is in principle tunable over a wide range
$10^{16}-10^{18}cm^{-3}$. Besides, the carriers do not freeze out
at low temperatures and have been shown to exhibit Shubnikov
de-Haas oscillations\cite{mynewpaper}.  Such degenerate 3DEGs
typically have a high density of dislocations ($N_{disl}\approx
10^{9}cm^{-2}$) and should be an ideal testing ground for our
theoretical predictions.

In summary, we derived the transport to quantum scattering times
for dislocation scattering for the three and two-dimensional
electron gases.  We compared the results to impurity scattering
and found the effect to be stronger for dislocation scattering for
both cases.  We attribute this to the inherent anisotropy in
scattering events due to the geometry of dislocations.  We point
out experimental systems where our results will prove helpful.







\begin{references}
\bibitem{dss} S. Das Sarma and F. Stern, Phys. Rev. B, {\bf 32}, 8442 (1985).
\bibitem{hsu_walu} L. Hsu and W. Walukiewicz, Appl. Phys. Lett., {\bf 80}, 2508 (2002).
\bibitem{gold} A. Gold, Phys. Rev. B., {\bf 38}, 10798 (1988).
\bibitem{look_prl} D. C. Look and J. R. Sizelove, Phys. Rev. Lett., {\bf 82}, 1237 (1999).
\bibitem{myfirstapl} D. Jena, A. C. Gossard, and U. K. Mishra, Appl. Phys. Lett., {\bf 76}, 1707 (2000).
\bibitem{elhamri} S. Elhamri et. al., Phys. Rev. B {\bf 57}, 1374 (1998).
\bibitem{arakawa} Z. W. Zheng et. al., Phys. Rev. B {\bf 62}, R7739 (2000).
\bibitem{frayssinet} E. Frayssinet et. al., Appl. Phys. Lett. {\bf 77}, 2551 (2000).
\bibitem{saxler} A. Saxler et. al. J. Appl. Phys. {\bf 87}, 369 (2000).
\bibitem{elhamriJAP} S. Elhamri et. al., J. Appl. Phys. {\bf 88}, 6583 (2000).
\bibitem{brana} A. F. Brana, C. Diaz-Paniagua, F. Batallan, J. A. Garrido, E. Munoz and F. Omnes, J. Appl. Phys. {\bf 88}, 932 (2000).
\bibitem{MyNewAPL} D. Jena et. al, Accepted for publication in Appl. Phys. Lett.
\bibitem{seeger} K. Seeger {\em Semiconductor Physics, an introduction} (Springer Verlag, Berlin 1999), p. 222.
\bibitem{podor} B. P\"{o}d\"{o}r, Phys. Status Solidi {\bf 16}, K167 (1966).
\bibitem{ambacher} O. Ambacher {\em et al.}, J. Appl. Phys. {\bf 87}, 334 (2000).
\bibitem{ibbo} J. P. Ibbetson {\em et. al.}, Appl. Phys. Lett. {\bf 77}, 250 (2000).
\bibitem{mynewpaper} D. Jena and U. K. Mishra, cond-mat/0209664
\end{references}
\end{document}